\documentstyle[12pt]{article}
\begin{document}

\centerline{\bf Irreversible Deposition of Line Segment Mixtures}
\centerline{\bf on a Square Lattice: Monte Carlo Study}
\centerline{$\;$}
\centerline{Jae Woo Lee\footnote{e-mail:jwlee@munhak.inha.ac.kr}}
\centerline{$\;$}
\noindent{Department of Physics, Inha University,
Inchon 402-751, KOREA}
\centerline{$\;$}
\noindent{Department of Physics, Clarkson University, 
Potsdam, NY 13699-5820, USA}

\begin{abstract}
We have studied kinetics of random sequential adsorption of mixtures on
a square lattice using Monte Carlo method. Mixtures of linear
 short segments
and long segments were deposited with the probability
$p$ and $1-p$, respectively. For fixed lengths of
each segment in the  mixture, the jamming limits decrease when $p$
increases. The jamming limits of mixtures always are greater than those
of the pure short- or long-segment deposition.
 For fixed $p$ and fixed length of
the short segments, the jamming limits have a maximum when the length
of the long segment increases. We conjectured a kinetic equation for
the jamming coverage based on the data fitting.
\end{abstract}


\newcommand{\beq}{\begin{equation}}
\newcommand{\eeq}{\end{equation}}
\newpage

Random sequential adsorption(RSA) is a model of irreversible
deposition of fixed-shape objects[1-3].
The objects are deposited randomly and
irreversibly on a substrate. When a depositing object overlaps
a deposited object,
 the depositing object is removed from the system and
another depositing is attempted sequentially. When we allow a
formation of a single layer, the substrate does not reach the full
coverage at the long time limit. The jamming coverage $\theta(t)$ is
defined by the total number of covered sites
 divided by the lattice size. The
coverage converges to a particular value, jamming
limit $\theta(\infty)$, in the long time limit.
RSA is realized in experimental studies of protein and colloid
particle adhesion on surfaces under the conditions of negligibly slow
surface relaxation[4-7].

The theoretical studies including 
rate equations[8-10], series expansion[11], and Monte Carlo
method[12-18]
were reported for both continuous and discrete
models. Exact solutions are also available for one-dimensional 
models[13, 14,19-21].
 In lattice deposition models the jamming coverage is asymptotically
exponential
\begin{equation}
\theta(t) = \theta(\infty) - A\exp(-Bt)
\end{equation}
where $A$ and $B$ are parameters which depend on the dimensionality of
the substrate and on the shape of the depositing objects.
There are several studies of kinetics of mixture depositions[22-28].
 Two
different kinds of objects are deposited on the substrate with a
different
adsorption probability of each object. Exact solutions were reported
for RSA of mixtures of monomer and linear $k$-mers in a one-dimensional
lattice[28]. They showed that the addition of pointlike particles 
modifies in a nonuniversal way the form of the long-time convergence law
of the approach to the jamming coverage. RSA of arbitrary mixtures of
line segments of two different lengths were solved analytically on the
one-dimensional lattice[29-30]. The deposition  on a square lattice of a
mixture of line segments 
of length $k_1$ and $k_2$ (end-on model), chosen with equal
probability, have been studied by \u{S}vraki\'{c} and Henkel using Monte Carlo
method[26]. They found two inequality relations for the jamming limits 
\begin{equation}
\theta(k_1, k_2, \infty) \ge \theta(k_1,\infty) \ge \theta(k_2,\infty)~~~~{\rm
for}~~~ k_2 \ge k_1
\end{equation}
and
\beq
\theta(k_1, k^{\prime}_2, \infty) \ge \theta(k_1, k_2, \infty) ~~~~{\rm
for } ~~~ k^{\prime}_2 \ge k_1
\eeq
where $\theta(k_1,k_2,\infty)$ is the jamming limit of the jamming
coverage $\theta(k_1,k_2,t)$ of a mixture and $\theta(k,\infty)$ is the
jamming limit of single $k$-mer.
They also proposed the time dependence of the jamming coverage
\beq
\theta(k_1,k_2,t) = \theta(k_1,k_2,\infty) - A(k_1,k_2) \exp(-
B(k_1,k_2) t)
\eeq
with $B(k_1,k_2) \simeq 1.0$ for $k_1 \neq k_2$. 

In the present work we have studied kinetics of RSA of mixture having a
general deposition probability of each segment, that is, the
deposition probability $p$ for short length of linear $k_1$-mers
and $1-p$  for long length of linear $k_2$-mers. Monte Carlo simulations
have been performed on a square lattice $L \times L$. We used
 lattices of
size $L=512$. We randomly select a lattice site and try to deposit a
line segment of length $k_1$ with the probability $p$ or 
length $k_2$ with the
probability $1-p$. We always take $k_1 \leq k_2$.  If a chosen site is
occupied by a deposited object, the attempt fails, the time is
increased by one unit, and a new site and a new line segment are
selected. If a chosen site is empty, then we randomly select a
direction of the four possible orientations. If all $k_1 -1$ (or $k_2
-1$) neighbor sites are empty, the selected sites are occupied. 
If any neighbor site
is occupied by previously deposited object in chosen direction, the
attempt fails and the time is increased by one unit and  we
 try a new site.
This model is standard model of RSA[25]. 
In end-on model, all possible
directions of a selected site are consumed by occuping 
selected line segments. 
After a long time, the system is close to the jamming limit. At late-time
stage we check all empty sites. If an empty site has at least one
possible direction which is available to accummulate 
a short line segment
$k_1$-mer,  then empty site is marked as an accessible site. If an
empty site has no possible direction to occupy $k_1$-mer, that site is
an inaccessible site. We further try to occupy line segments 
at accessible sites. If there are no accessible sites, the system is in
the jamming limit. We always use periodic boundary conditions. One
Monte Carlo time step is 
defined by the total number of attempts to select a
site divided by the total number of lattice sites. The data are
averaged over 50 independent runs for each choice of mixtures and each
deposition probability.

We present the typical jamming configuration of mixtures $(k_1 = 4,
k_2=16)$ in Fig. 1 (a) and $(k_1=4, k_2=32)$ in Fig. 1 (b) for the short
$k_1$-mer deposition probability $p=0.5$. The lattice size is $128
\times 128$ square lattice with periodic boundary conditions.  The
configurational structure of jamming limit has a lot of local
structures which have parallel short and long line segments.
In Fig. 2 we plot $\ln[ \theta(k_1,k_2,\infty) - \theta(k_1,k_2,t) ] $
versus time for $k_1 =2$ and $p=0.1$. At long time limit the lines are
linear and parallel. In  such a standard model the jamming coverages
follow the exponential behaviour like eq. (4). We have calculated the
slopes of the parallel lines for various length of $(k_1, k_2)$ pairs
and $p$.  We concluded $B(k_1,k_2,p)=p/2$ from linear slopes. In Fig. 2
we can observe the independence of $B(k_1,k_2,p)$ on the length of 
each line segment. All lines are parallel and the slopes depend only
on the deposition probability. These results are different from the
observation of end-on model $B(k_1,k_2,p=1/2) =1$[26] and single 
adsorption model $B(k)=2$[31]. In end-on model we observed
$B(k_1,k_2,p)=2p$[32].

In Fig.3 we show the jamming limits versus the short $k_1$-mer
deposition probability $p$ for $(k_1=2, k_2)$ mixtures. The jamming
limits decrease monotonically when $p$ increases.
The jamming limit approaches to a
value of $\theta(k=2,\infty)=0.9068$ at $p \rightarrow 1$. 
At $p \rightarrow 0$ the jamming limit shows a singularity.
For example, the jamming limit drops to a value of
pure single segment deposition
$\theta(k_1=2,k_2=3,\infty)=0.8467$ at $p=0$.
This singularity
was already observed in one-dimensional adsorption of mixtures[30].
For fixed length of $k_1$ and $k_2$ we observed that the jamming limit
satisfied the same inequality  relation as eq. (2). The mixture
depositions lead to more efficient jamming coverage than depositions
of the single line segment.
The jamming coverages are controlled by the long segment
 of mixtures up to cross-over times $t_{\times}$. At $t > t_{\times}$.
The local empty spaces are shorter than the length of the long segments.
Therefore, the short segments are further adsorbed on the substrate. 

In Fig.4 we plot the jamming limit $\theta(k_1=2, k_2, \infty;p)$
versus the length of a long segment for fixed length of short segment
$k_1=2$ and various  $p$-values. For a fixed value  $k_2$, the jamming
limits satisfy $\theta(k_1,k_2,\infty;p_2) >
\theta(k_1,k_2,\infty;p_1)$ when $p_2 > p_1$. For a fixed value $p$ the jamming
limits reach a maximum value at a particular value $k_2^*$. A
combination $(k_1, k_2^*)$ of mixtures at a fixed value of $p$ gives
the most efficient jamming coverage at long time limit. We check the
finite size effects for $p=0.5$ using $256 \times 256$ and $1024 \times
1024$ lattices. We observed that the effects of finite size are
negligible. We concluded that these maximum behaviours are intrinsic in
mixture deposition on such a standard model.
 We also observed the maximum
behaviour for end-on model at a certain range of the probability
$p$. (These results will be published elsewhere[32].) When the length of
the long segments increases for fixed length of $k_1$, the adsorbed long
segments induce that the newly adsorbing long segments
locally have the same direction as adsorbed one as shown in Fig. 1 (b).
Among the absorbed parallel long segments, the short $k_1$-mers further
adsorbed to the direction of the long $k_2$-mers. These adsorptions
induced local one-dimensional characteristics of the short segments and
the jamming limits decrease for $k_2 > k_2^*$. 

We have performed the data fitting to check the parameter dependence of
the amplitude $A(k_1, k_2, p)$.  We concluded that the amplitude
has the functional form as 
$A(k_1, k_2,p)=C_o (1/k_1)^2 \exp[ C_1 (1-p) + C_2 (1-p)/k_2]$. The
amplitudes are well fitted for the choice of the constant $C_o=0.8 \pm
0.1$, $C_1 = -0.5 \pm 0.2 $ and $C_2 = -1.3 \pm 0.1$ for all $k_1 \leq
k_2$ mixture combinations[32].

In summary we observed that the mixture depositions cover the substrate
more efficiently than the deposition of single-length segments.
We observed that the jamming limits show a
maximum behaviour for fixed $k_1$ and $p$ when the lengths of $k_2$
vary. The jamming coverage shows exponential behaviour. We proposed
a functional form of the amplitude of coverage from the data fitting.

{\sl This work has been supported by Inha University and by the Basic
Science Institute Program, Ministry of Education, Project No.
BSRI-97-2430. I wish to thank Professor Vladimir Privman for his
critical reading of this manuscript.
}

\newpage
\newcommand{\jpa}{J. Phys. A}
\newcommand{\pra}{Phys. Rev. A}
\newcommand{\prb}{Phys. Rev. B}
\newcommand{\prl}{Phys. Rev. Lett.}
\newcommand{\pre}{Phys. Rev. E}
\newcommand{\jcp}{J. Chem. Phys.}
\centerline{\bf References}

\ 

\frenchspacing{

\noindent\hang [1] J. W. Evans, Rev. Mod. Phys. {\bf 65}, 1281(1993).

\noindent\hang [2] M. C. Bartelt and V. Privman, Int. J. Mod. Phys. B
{\bf 5}, 2883(1991).

\noindent\hang [3] V. Privman, Trends Stat. Phys. {\bf 1}, 89(1994).

\noindent\hang [4] P. J. Flory, J. Am. Chem. Soc. {\bf 61}, 1518(1939).

\noindent\hang [5] J. Feder and I. Giaever, J. Colloid Interf. Sci.
{\bf 78}, 144(1980).

\noindent\hang [6] G. Y. Onoda and E. G. Linger, \pra {\bf 33},
715(1986).

\noindent\hang [7] V. Privman, N. Kallay, M. F. Haque and 
E. Matijevi\'{c},
J. Adhesion Sci. Technol. {\bf 4}, 221(1990).

\noindent\hang [8] J. Evans and R. S. Nord, J. Stat. Phys. {\bf 38},
681(1985).

\noindent\hang [9] P. Schaaf, J. Talbot, H. M. Rabeony and H. Reiss,
\jcp {\bf 92}, 4826(1988).

\noindent\hang [10] P. Schaaf and J. Talbot, \prl {\bf 62}, 175(1989).

\noindent\hang [11] R. D. Vigil and R. M. Ziff, \jcp {\bf 91},
2599(1989).

\noindent\hang [12] M. Nakamura, \pra {\bf 36}, 2384(1987).

\noindent\hang [13] P. Nielaba, V. Privman and J. S. Wang, \jpa {\bf
23}, L1187(1990).

\noindent\hang [14] P. Nielaba, V. Privman and J. S. Wang, \prb {\bf 43},
3366(1991).

\noindent\hang [15] J. D. Sherwood, \jpa {\bf 23}, 2827(1990).

\noindent\hang [16] R. M Ziff and R. D. Vigil, \jpa {\bf 23},
5103(1990).

\noindent\hang [17]  M. C. Bartelt and V. Privman, \jcp {\bf 93},
6820(1990).

\noindent\hang [18] B. Bonnier, M. Hontebeyrie, Y. Leroyer, C. Meyers
and E. Pommiers, \pre {\bf49}, 305(1994).

\noindent\hang [19] B. Widom, \jcp {\bf 44}, 3888(1966).

\noindent\hang [20] R. Pomeau, \jpa {\bf 13}, L193(1980).
 
\noindent\hang [21] R. Swendsen, \pra {\bf 24}, 504(1981).

\noindent\hang [22] B. Mellin and E. E. Mola, J. Math. Phys. {\bf 26},
514(1985).

\noindent\hang [23] B. Mellin, J. Math. Phys. {\bf 26}, 1769,(1985),
{\bf 26}, 2930(1985).

\noindent\hang [24] G. C. Barker and M. J. Grimson, Mol. Phys. {\bf
63}, 145(1988).

\noindent\hang [25] R. S. Nord and J. W. Evans, \jcp {\bf 93},
8397(1990).

\noindent\hang [26] N. M. \u{S}vraki\'{c}
 and M. Henkel, J. Phys. I {\bf 1}, 791(1991).

\noindent\hang [27] Lj. Budinski-Petlovi\'{c} 
and U. Kozmidis-Laburi\'{c},
Physica A {\bf 236}, 211(1997).

\noindent\hang [28] M. C. Bartelt and V. Privman, \pra {\bf 44} R2227(1991).

\noindent\hang [29] G. J. Rodgers, \pra {\bf 45}, 3443(1992).

\noindent\hang [30] B. Bonnier, Europhy. Lett. {\bf 18}, 297(1992).

\noindent\hang [31] S. S. Manna and N. M. \u{S}vraki\'{c}, \jpa {\bf 24},
L671(1991).

\noindent\hang [32] J. W. Lee, {\em in preparation}

}

\newpage
\centerline{\bf Figure Captions}

\ 

\noindent\hang Figure 1: Typical configurations of
mixture deposition on $128 \times 128$ square lattices for (a) $k_1=4,
k_2=16$ and (b) $k_1=4, k_2=32$.

\ 

\noindent\hang Figure 2: The plot of $\ln [ \theta(k_1, k_2,\infty) -
\theta(k_1, k_2,t) ]$ versus time for $k_1=2$, $p=0.1$ and 
$k_2$=4, 8, 16, 32, 64 from bottom to top.

\ 

\noindent\hang Figure 3: The jamming limits versus the short segment
deposition probability $p$ for $k_1=2$ and $k_2=3(\diamond)$, $4(+)$,
$8(*)$, $16(\triangle)$, $32(\Box)$, and $64(\times)$.

\ 

\noindent\hang Figure 4: The jamming limits $\theta(k_1=2,
k_2,\infty;p)$ versus the length of the long segment for a fixed length
of the short segment $k_1=2$ and $p=0.1(\triangle)$, $0.3(\times)$,
$0.5(\Box)$, $0.7(+)$, and $0.9(\diamond)$.

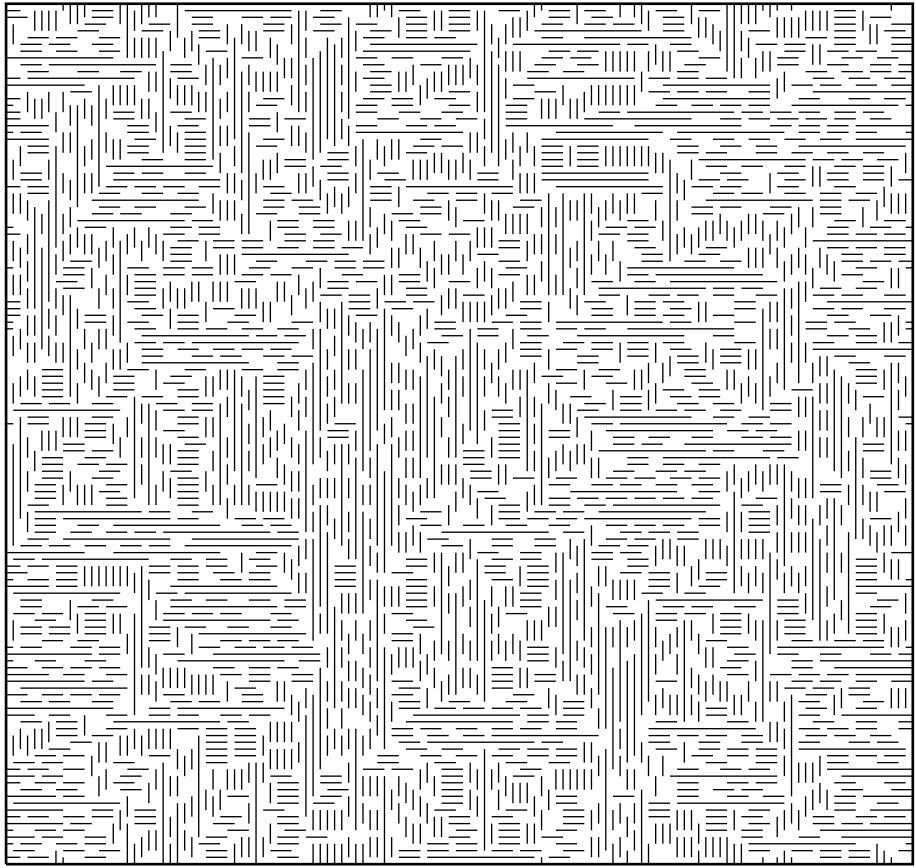
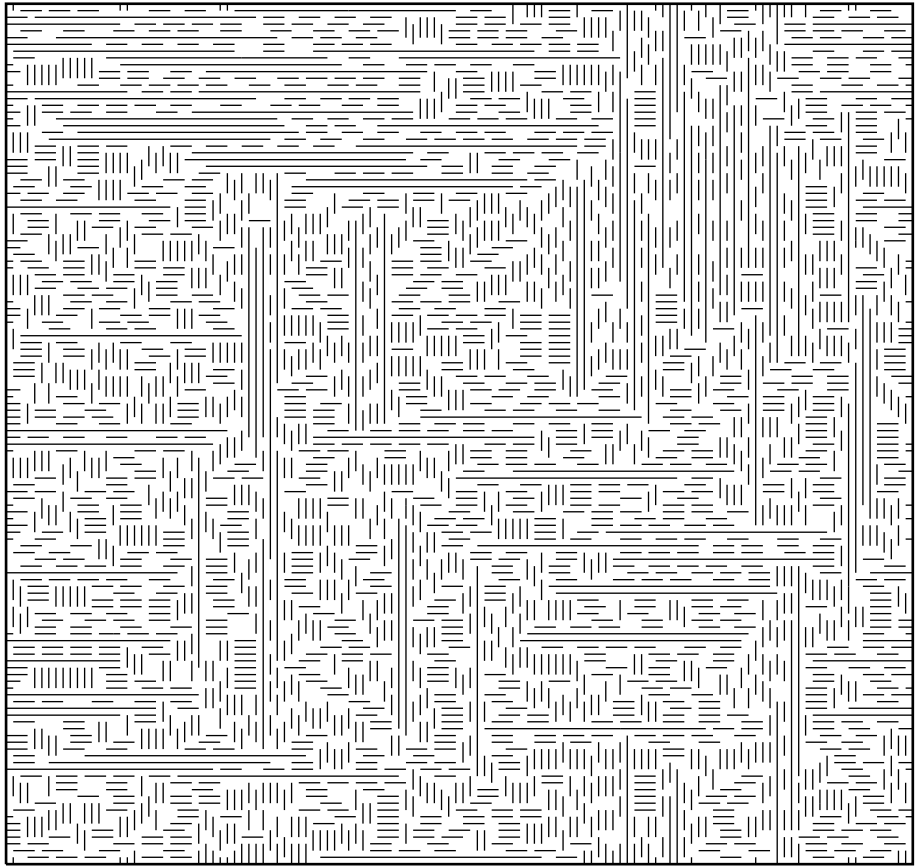
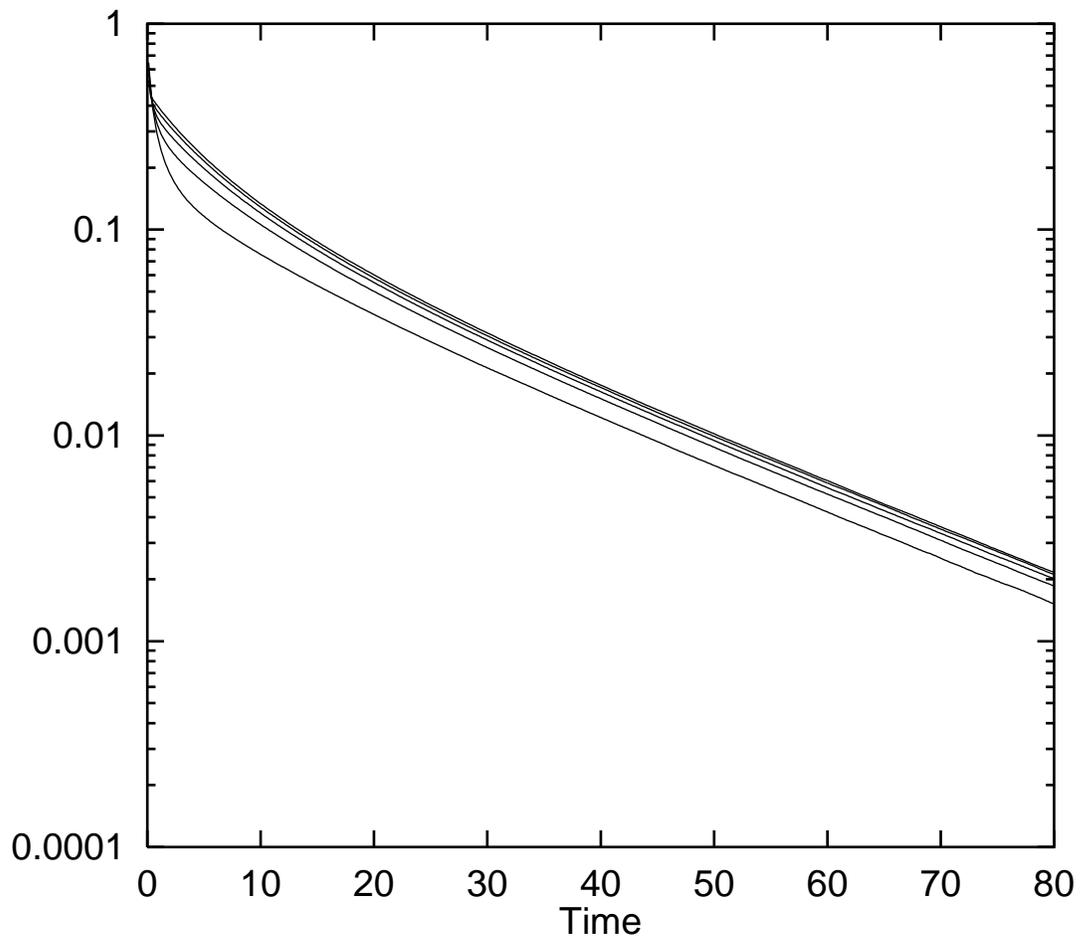
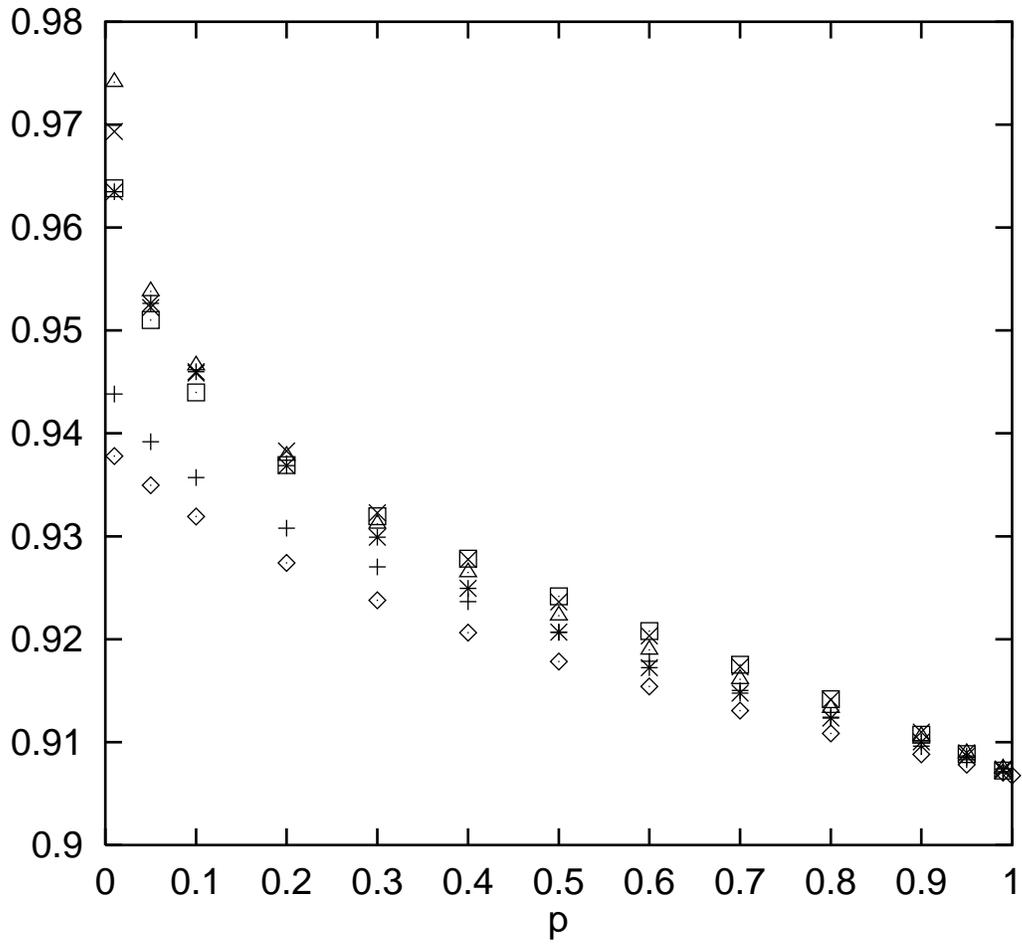
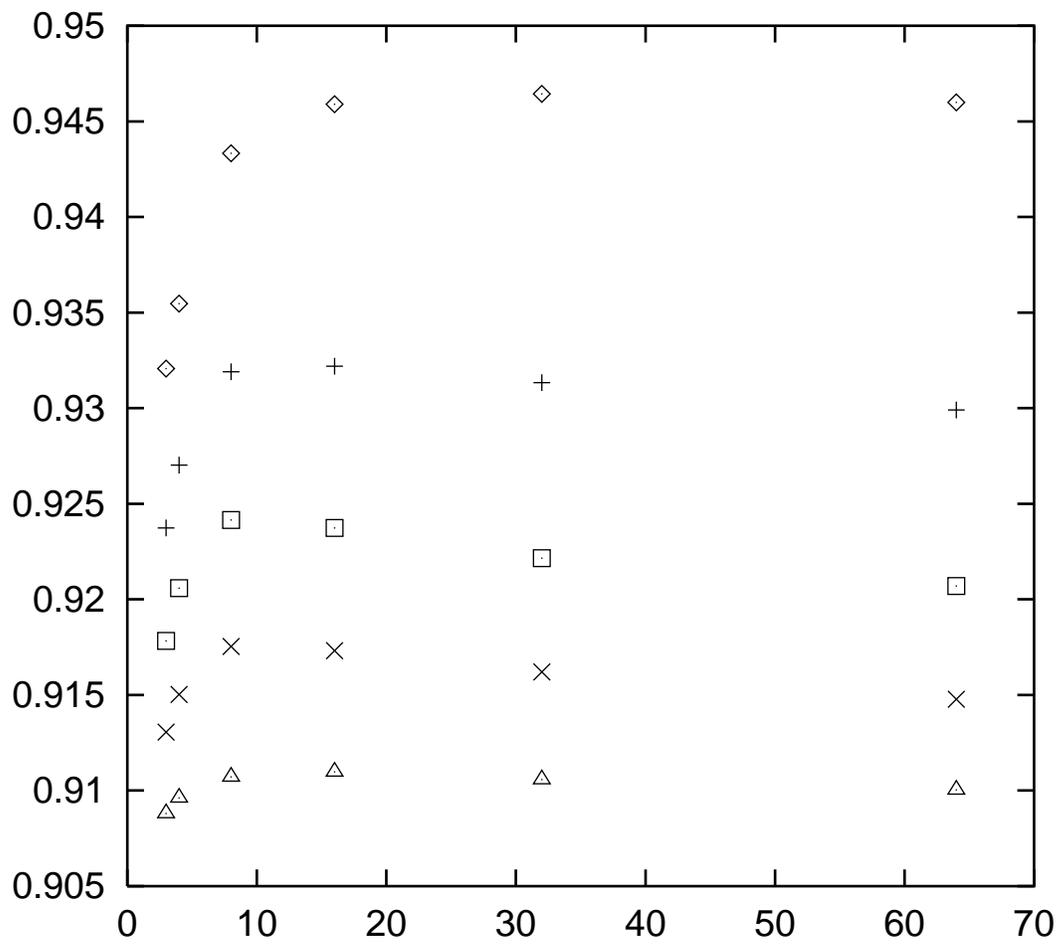
\end{document}